\documentclass[]{raa} 

\usepackage{graphicx,times,epsf}             

\begin{document}

   \title{Synchrotron X-ray diagnostics of cutoff shape of nonthermal 
          electron spectrum at young supernova remnants
}

   \volnopage{Vol.14 (2014) No.2, 165--178}      
   \setcounter{page}{1}          

   \author{Ryo~Yamazaki
      \inst{1}
   \and Yutaka~Ohira
      \inst{1}
   \and Makoto~Sawada
      \inst{1},
   \and Aya~Bamba
      \inst{1}
   }

   \institute{Department of Physics and Mathematics, Aoyama Gakuin University, 
              5-10-1 Fuchinobe, Sagamihara 252-5258, Japan;
             {\it ryo@phys.aoyama.ac.jp}\\
   }

   \date{Received~~2009 month day; accepted~~2009~~month day}

\abstract{ 
Synchrotron X-rays can be a useful tool to investigate 
electron acceleration at young supernova remnants (SNRs).
At present, since the magnetic field configuration around the shocks
of SNRs is uncertain, it is not clear whether electron
acceleration is limited by SNR age, synchrotron cooling, or
even escape from the acceleration region.
We study whether the acceleration mechanism can be constrained by
the cutoff shape of the electron spectrum around the maximum energy.
We derive analytical formulae of the cutoff shape in each case
where the maximum electron energy is determined by
SNR age, synchrotron cooling and escape from the shock.
They are related to the energy dependence of the electron diffusion coefficient.
Next, we discuss whether information on the cutoff shape can be provided
by observations in the near future which will simply give
the photon indices and the flux ratios in the soft and hard X-ray bands. 
We find that if the power-law index of the electron spectrum is independently determined
by other observations, then we can constrain the cutoff shape by comparing
theoretical predictions of the photon indices and/or the flux ratios 
with observed data which will be measured by {\it NuSTAR} and/or {\it ASTRO-H}.
Such study is helpful in understanding the acceleration mechanism.
In particular, it will supply another
independent constraint on the magnetic field strength around
the shocks of SNRs.
}

   \authorrunning{R. Yamazaki et al.}            
   \titlerunning{Electron cutoff shape}  

   \maketitle

%
%
\section{Introduction}           
\label{sec:intro}

Observations of synchrotron X-rays indicate that 
young supernova remnants (SNRs) act as electron accelerator up to 10--100~TeV
(e.g., Koyama et al.~\cite{koyama95}).
Spectral fitting of the synchrotron radiation from radio to X-ray bands
gives us information on the acceleration mechanism.
For example, the maximum energy of electrons $E_{\rm max,e}$
and the magnetic field strength $B$ are constrained by 
measuring the roll-off frequency of the spectrum, $\nu_{\rm roll}$,
which is proportional to $BE_{\rm max,e}{}^2$
(Reynolds~\cite{reynolds98}; Reynolds \& Keohane~\cite{reynolds99}).
Furthermore, if the rapid cooling time of 1--10~yr is responsible for the
narrowness of thin filaments close to the shock front  
(Vink \& Laming~\cite{vink03}; Bamba et al.~\cite{bamba03b},
\cite{bamba05a}, \cite{bamba05b}; Yamazaki et al.~\cite{yamazaki04})
and/or time variability of synchrotron X-rays
(Uchiyama et al.~\cite{uchiyama07}; Uchiyama \& Aharonian~\cite{uchiyama08}),
then $B\sim0.1$--1~mG is inferred, so that $E_{\rm max,e}\sim10$~TeV.
Given the values of $E_{\rm max,e}$ and $B$ as well as
the SNR age $t_{\rm age}$,
we can even estimate the maximum energy of protons $E_{\rm max,p}$,
which is proportional
to $E_{\rm max,e}{}^2B^2t_{\rm age}$ (see \S~\ref{subsec:Emax}).

For young SNRs, $\nu_{\rm roll}$ is typically below the X-ray band 
(e.g., Bamba et al.~\cite{bamba03b}, \cite{bamba05a}, \cite{bamba05b}),
so that synchrotron X-rays are emitted by electrons whose energy
is near $E_{\rm max,e}$.
Hence it is expected that information on the electron spectrum
at highest energy range can be extracted from the soft and hard X-ray 
data. In particular, the recently launched satellite
{\it The Nuclear Spectroscopic Telescope Array (NuSTAR)} 
(Hailey et al.~\cite{hailey10}; Harrison et al.~\cite{harrison2013})
 and 
{\it ASTRO-H} (Takahashi et al.~\cite{takahashi10})
that will start in the near future
 will observe hard X-rays whose photon energy 
is larger than 10~keV.
In this paper, we focus on the cutoff shape of the electron spectrum
and show that it may provide another independent way to tackle the
problems of particle acceleration at young SNRs (\S~\ref{sec:cutoff}).
Then, we show that observations by these satellites will
play an important role
in this context (\S~\ref{sec:synch} and \ref{sec:discuss}).


\section{Cutoff shape of the electron spectrum}
\label{sec:cutoff}

We assume the shape of the electron spectrum
around the maximum energy, $E_{\rm max,e}$, in the simple form,
\begin{equation}
N(E)\propto E^{-p} \exp\left[-\left(E/E_{\rm max,e}\right)^a\right]~~.
\label{eq:spec}
\end{equation}
As discussed in the following, the spectral index $p$ and 
the cutoff shape parameter $a$ contain
rich information on the acceleration mechanism.
We parameterize the functional form of the electron spectrum, Equation~(1),
to be as simple as possible so that
the result can be compared with observations.
In practice, when we consider X-ray synchrotron emission from
typical young SNRs, only the energy range around $E_{\rm max,e}$ is
important because the roll-off energy of the synchrotron radiation is
much smaller than 2 keV, so that detailed fluctuations of the 
electron spectrum caused by more realistic but somewhat complicated
models, such as the nonlinear model, may not be important.

\subsection{Spectral index $p$}
\label{sec:index}

When the diffusive shock acceleration works 
at the adiabatic shock with a compression
ratio of $r$ and  the energy loss effects are negligible,
the spectral index is given by $p=(r+2)/(r-1)$ 
in the test-particle limit 
(Bell~\cite{bell78}; Blandford \& Ostriker~\cite{blandford78}).
In particular, $p=2.0$ in the case of the strong shock limit $r=4$.
This simplest case may not be applicable for actual young SNRs, in which
GeV-to-TeV gamma-ray observations of youngest SNRs such as Cas~A and
Tycho suggest $p>2.0$ 
(Abdo et al.~\cite{abdo10casA}; Giordano et al.~\cite{giordano12}).
This fact has already been inferred from both the radio synchrotron spectrum
and the propagation model of cosmic rays 
(e.g., Strong \& Moskalenko~\cite{strong98}; Putze et al.~\cite{putze09};
Shibata et al.~\cite{shibata11}).

If the magnetic field is strong enough (e.g., in the Bohm diffusion
model, $K(E)\propto E$, see equation~(\ref{eq:cond_Eb})),
 synchrotron cooling is responsible for 
a spectral break, above which the electron spectrum becomes 
softer and the spectral index increases by 1.0, so that
$p>3.0$ (e.g., Longair~\cite{longair94}).
Note that the spectral steepening by one power occurs for
homogeneous stationary sources.
It is also noted that sometimes the synchrotron cooling effect causes
spectral hardening or pile-up
(Longair~\cite{longair94}; Drury et al.~\cite{drury99};
Zirakashvili \& Aharonian~\cite{zirakashvili07}).
For typical parameters of young SNRs, however,
such a hard component can hardly be seen for more than a decade of electron energy.
At most a small bump is formed just below $E_{\rm max,e}$,
which may be seen as a small excess in the radiation spectrum
(see \S~\ref{sec:discuss}).

\subsection{Cutoff shape parameter $a$}
\label{sec:cutoff_shape}

The cutoff shape parameter $a$ also depends on the details of 
electron acceleration, such as the magnetic field strength
and the energy dependence of the diffusion coefficient.
We assume that the diffusion coefficient of high-energy
electrons has the following power-law form
\begin{equation}
K(E)\propto E^\beta~~.
\label{eq:diff}
\end{equation}
Usually the Bohm diffusion, $\beta=1$, is widely adopted.
However, in general $\beta$ may deviate from unity.
For example, it is well known that $\beta$ becomes 1/3 
if particle diffusion is considered in the Kolmogorov
magnetic turbulence, but $\beta=1/2$ for Kraichnan turbulence
(e.g., Blandford \& Eichler~\cite{blandford87})\footnote{
In general, if the spectrum of magnetic turbulence has a form 
$E_k\propto k^{-s}$, then $\beta$ and $s$ are related 
as $\beta=2-s$ under the assumption that accelerated particles scatter
via wave-particle resonance interaction.
In particular, for Kolmogorov ($s=5/3$) and
Kraichnan ($s=3/2$) turbulence, we obtain
$\beta=1/3$ and 1/2, respectively.
}.
So in these cases, the value of $\beta$ even tells us the
properties of magnetic turbulence.
In another context, if the wave damping due to ion-neutral
collisions is significant, $\beta$ may approach $\approx2$
(Bykov et al.~\cite{bykov00}; Lee et al.~\cite{lee12}).

In the cooling limited case, where $E_{\rm max,e}$ is determined by
the balance of acceleration and synchrotron cooling
($t_{\rm acc}(E)=t_{\rm syn}(E)$), the spectral shape factor
$a$ is related to $\beta$ as
\begin{equation}
a = \beta+1 ~~,
\label{eq:shape_cooling}
\end{equation}
which is analytically derived as in \S~\ref{sec:appendix2}.
In the age-limited case,
where the synchrotron cooling effect is neglected and
$E_{\rm max,e}$ is determined by the finite age,
Kang et al.~(\cite{kang09}) have given 
\begin{equation}
a = 2\beta ~~,
\label{eq:shape_age}
\end{equation}
by fitting their results of numerical simulation
(see also Kato \& Takahara~\cite{kato03}).
Note that in the case of Bohm diffusion (that is, $\beta=1$),
both age-limited and cooling-limited cases give the same value, $a=2$.
If $\beta\ne1$, the values of $a$ for the two cases are different.

It may happen that the maximum energy is limited by the escape process
(Ptuskin \& Zirakashvili~\cite{ptuskin05}; Drury et al.~\cite{drury09};
Caprioli et al.~\cite{caprioli09}; Ohira et al.~\cite{ohira10}).
Here we consider the simplest case (see \S~\ref{sec:escape}), in which
a free escape boundary exists upstream of
the shock front. In the test particle limit,
we analytically derive
\begin{equation}
a = \beta~~.
\label{eq:shape_escape}
\end{equation}
If nonlinear effects, in particular the decay of self-excited upstream turbulence,
are taken into account, then $a$ may be slightly larger 
(e.g., Lee et al.~\cite{lee12}) although the
precise cutoff shape is at present highly uncertain 
(Ellison \& Bykov~\cite{ellison2011}).
However, as long as $\beta\approx1$, 
the escape process does not affect the maximum electron energy 
(Ohira et al.~\cite{ohira12})
and hence the cutoff shape  for young SNRs.

Equations~(3), (4), and (5), are derived on the assumption that
the shock velocity is constant with time, but the shock velocity of 
real SNRs decreases with time after the free expansion phase.
Some young SNRs are still in the free expansion phase or have just entered
the Sedov phase, so that the effect of shock deceleration is not important.
Even if the shock velocity is decreasing,
our assumption that the shock is stationary may not significantly 
influence the result because of the following reasons. 
In the cooling limited and escape limited cases, the acceleration time
of electrons with arbitrary energy is smaller than the dynamical time,
so that the shock velocity can be treated as stationary.
Even in the age-limited case,
high-energy particles produced around the shock suffer adiabatic expansion after
they are transported downstream of the shock and lose their energy.
Hence, at a given epoch, the spectrum for high-energy particles is
dominated by those which are being accelerated at that time, in other
words, the energy spectrum of particles does not depend so much 
on the past acceleration history. 
In particular, we are now interested
in the energy region near the upper end of the spectrum because as
seen in the following, X-rays are produced by
particles with energy near the maximum energy. 
In this energy regime, the electron spectrum
is well approximated by that for the stationary shock case.

So far, we have discussed the cutoff shape in the diffusion approximation for 
particle motion.
For the cooling limited and the age limited cases, the diffusion
approximation is valid even around the maximum energy,
because their gyro radii are much shorter than
the size of the acceleration region and/or escape boundary.
For the escape-limited case, the mean free path of a particle
becomes large around the escape boundary,
so that the diffusion approximation is invalid around the escape boundary
and the cutoff shape could be modified.
Even so, it can be said that the cutoff shape contains rich information
on the particle acceleration.


\section{Synchrotron X-ray diagnostics of cutoff shape}
\label{sec:synch}

In this section, we consider the synchrotron radiation from
electrons whose energy distribution is given by equation~(\ref{eq:spec}).
Assuming an isotropic pitch-angle distribution of electrons,
the energy spectrum
$F_\nu$~[erg~s$^{-1}$cm$^{-2}$Hz$^{-1}$]
of the synchrotron radiation is computed.
The energy flux, in the photon energy range between
$\varepsilon_1=h\nu_1$ and $\varepsilon_2=h\nu_2$ ($\nu_1<\nu_2$),
is given by
$F(\varepsilon_1-\varepsilon_2)=\int_{\nu_1}^{\nu_2}F_\nu d\nu$~.
We calculate photon indices and flux ratios 
in various energy bands,
in which we focus in the following.
These spectral quantities are determined if we specify four parameters, $a$, $p$, 
$E_{\rm max,e}$ and the magnetic field strength $B$ in the emitting region.
Note that the flux normalization is not necessary.
The field strength $B$ is only required to determine the frequency
that gives the peak of the $\nu F_\nu$
for the synchrotron radiation, a so called roll-off frequency 
$\nu_{\rm roll}(\propto BE_{\rm max,e}{}^2)$, which roughly 
corresponds to the characteristic synchrotron frequency of electrons with
$E_{\rm max,e}$ 
(Reynolds~\cite{reynolds98}; Reynolds \& Keohane~\cite{reynolds99}).
Different parameter sets but with the same value of $BE_{\rm max,e}{}^2$ give
the same photon indices and flux ratios.
Therefore, independent parameters are 
$a$, $p$ and $BE_{\rm max,e}{}^2$.

\begin{figure*}[t]
\includegraphics[bb=50 160 590 640,width=15cm,clip]{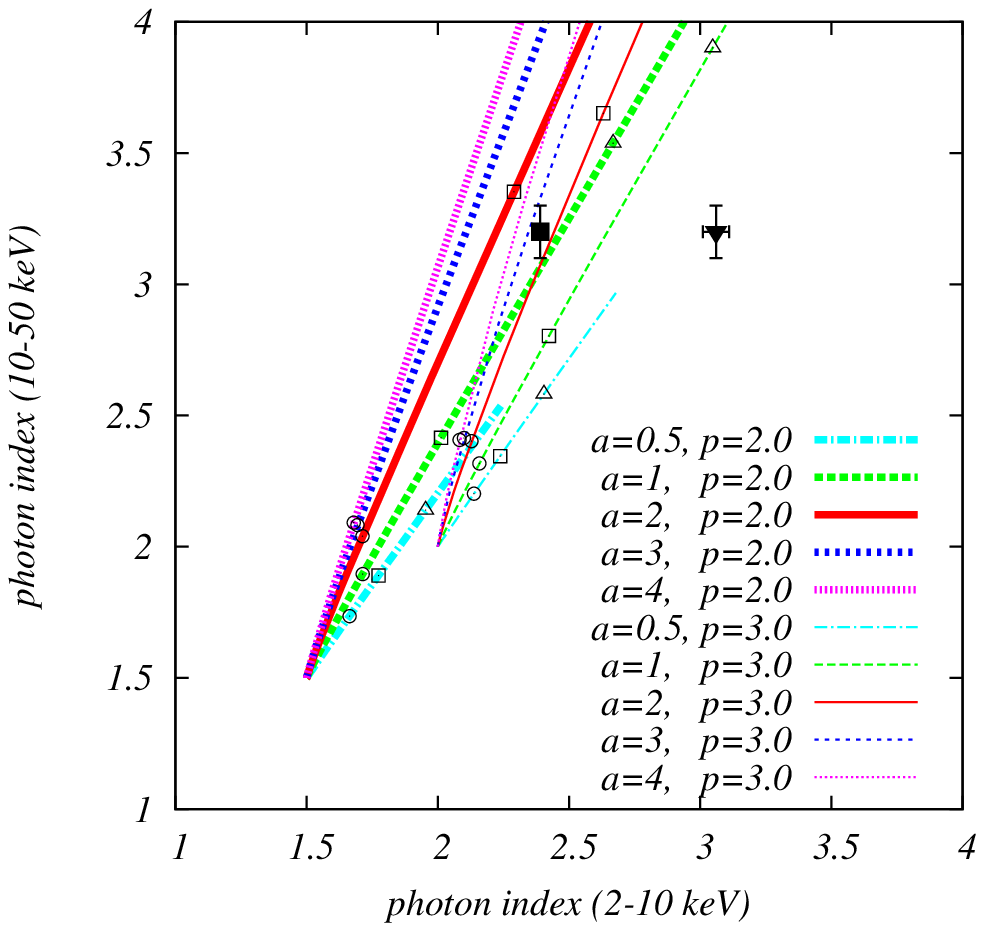}
\caption{ 
Photon index in the 10--50~keV band 
as a function of the 2--10~keV photon index. 
Along each line, both $a$ and $p$ are constant, and the quantity
$BE_{\rm max,e}{}^2$ changes from $10^3\mu{\rm G}~({\rm TeV})^2$ to $\infty$;
open triangles, squares, and circles are for 
$BE_{\rm max,e}{}^2=10^4$, $10^5$ and $10^6\mu{\rm G}~({\rm TeV})^2$,
respectively.
Thick lines are for $p=2.0$, while thin lines are for $p=3.0$.
Light blue, green, red, blue and purple lines are for
$a=0.5$, 1, 2, 3 and 4, respectively. 
The filled square and triangle are observed data for
RX~J1713.7$-$3946 and Cas~A, respectively (see section~4.1).
}
\label{fig1}
\end{figure*}

Figure~\ref{fig1} shows the
10--50~keV photon index as a function of
2--10~keV photon index. 
We adopt $p=2.0$ (thick lines) and 3.0 (thin lines),
and $a=0.5$ (light blue), 1 (green), 2 (red), 3 (blue) and 4 (purple).
Along each line, both $a$ and $p$ are constant, and the quantity
$BE_{\rm max,e}{}^2$ changes.
In the limiting case of $BE_{\rm max,e}{}^2\rightarrow\infty$,
rolloff frequency $\nu_{\rm roll}$ goes beyond the observation band,
which implies the synchrotron spectrum is well approximated by
$F_\nu\propto\nu^{-(p-1)/2}$.
Then the photon index becomes an asymptotic value,
\begin{equation}
\Gamma_\infty=\frac{p+1}{2}~~,
\end{equation}
so that $\Gamma_\infty=1.5$ and 2.0 for $p=2.0$ and 3.0, respectively,
which correspond to the left end of each line.
When $BE_{\rm max,e}{}^2$ decreases, the rolloff frequency $\nu_{\rm roll}$
crosses the observation bands, so the photon indices become larger than the asymptotic
value $\Gamma_\infty$.
After passing through the harder band 10--50~keV, $\nu_{\rm roll}$
crosses the softer 2--10~keV band, so the photon index in the former band
is larger than the latter one.
For fixed $p$, if $a$ becomes larger, the flux beyond $\nu_{\rm roll}$
decreases more rapidly, resulting in a larger photon index.
Hence each line has a steeper slope for larger $a$.
The value of $BE_{\rm max,e}{}^2$
is reduced to $10^3\mu{\rm G}~({\rm TeV})^2$, which corresponds to
the characteristic frequency of the synchrotron radiation of 
$2.9\times10^{16}$~Hz.
Light blue lines in all the Figures as well as the thick green line
in Figure~\ref{fig4} have a right end, which
corresponds to this lower limit.

\begin{figure*}[t]
\includegraphics[bb=10 160 550 640,width=15cm,clip]{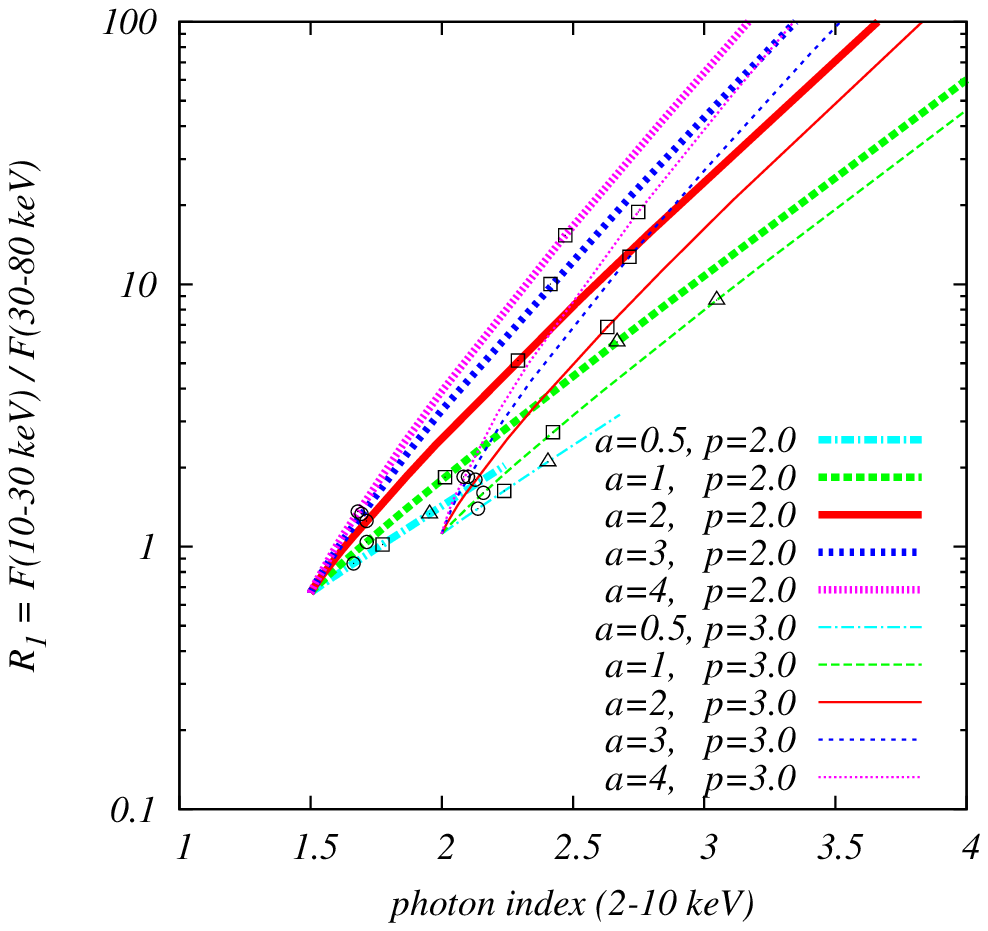}
\caption{ 
The flux ratio $R_1=F(10-30~{\rm keV})/F(30-80~{\rm keV})$
as a function of the 2--10~keV photon index. 
The meanings of each line and point are the same as in Fig.~1.
}
\label{fig2}
\end{figure*}

Figure~\ref{fig2} shows the flux ratio
$R_1=F(10-30~{\rm keV})/F(30-80~{\rm keV})$ as a function of
the 2--10~keV photon index.
In the case of $BE_{\rm max,e}{}^2\rightarrow\infty$,
the flux ratio
$R=F(\varepsilon_1-\varepsilon_2)/F(\varepsilon_3-\varepsilon_4)$
of two energy bands $\varepsilon_1-\varepsilon_2$~[keV]
and  $\varepsilon_3-\varepsilon_4$~[keV] has
an asymptotic value
\begin{equation}
R\rightarrow
\frac{\varepsilon_2^{2-\Gamma_\infty}-\varepsilon_1^{2-\Gamma_\infty}}
{\varepsilon_4^{2-\Gamma_\infty}-\varepsilon_3^{2-\Gamma_\infty}}~~,
\end{equation}
for $\Gamma_\infty\ne2$, but
\begin{equation}
R\rightarrow\frac{\ln(\varepsilon_2/\varepsilon_1)}
{\ln(\varepsilon_4/\varepsilon_3)}~~,
\end{equation}
for $\Gamma_\infty=2$.
Hence in the present case
($\varepsilon_1=10$~keV, $\varepsilon_2=\varepsilon_3=30$~keV
and $\varepsilon_4=80$~keV), we have
$\Gamma_\infty=1.5$ and $R_1\rightarrow0.668$ for $p=2.0$,
but $\Gamma_\infty=2.0$ and $R_1\rightarrow1.12$ for $p=3.0$.
When $BE_{\rm max,e}{}^2$ becomes small, both the flux
ratio and the 2--10~keV photon index become large,
however the decay slope is steeper for large $a$.

\begin{figure*}[t]
\includegraphics[bb=50 160 590 640,width=15cm,clip]{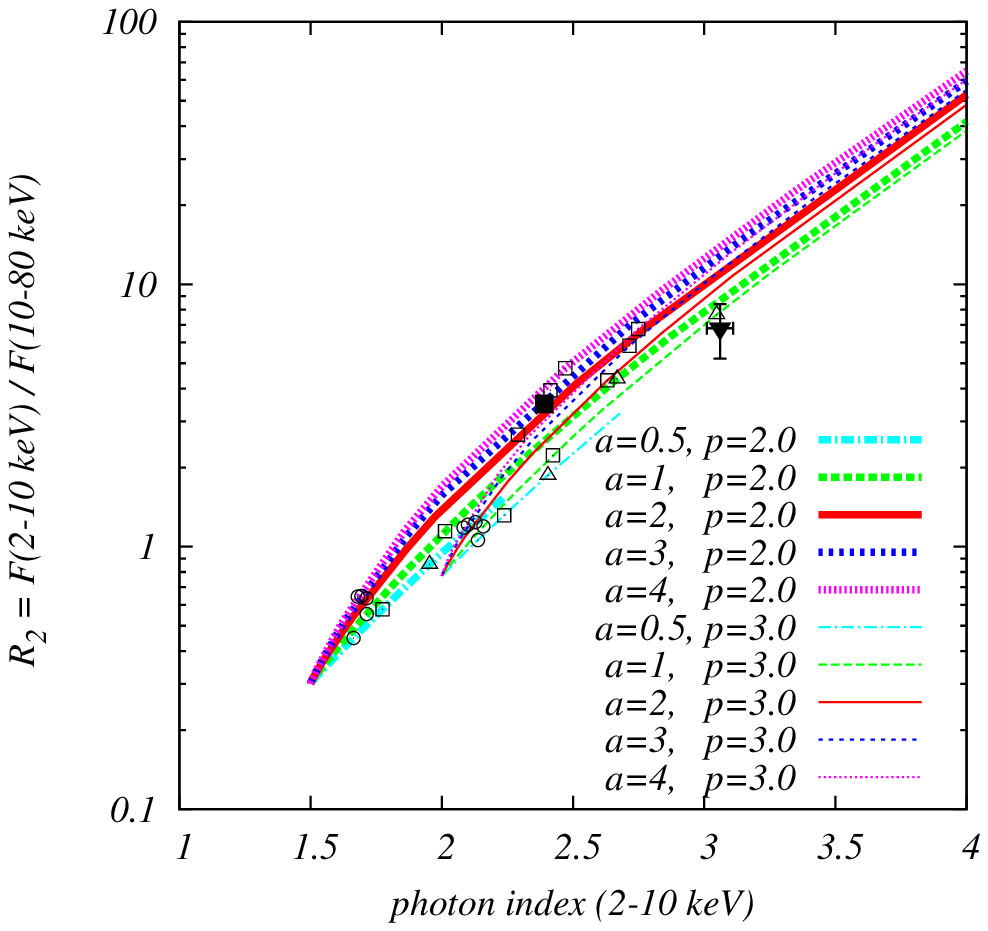}
\caption{The flux ratio
$R_2=F(2-10~{\rm keV})/F(10-80~{\rm keV})$ as a function of
the 2--10~keV photon index.
The meanings of each line and point are the same as in Figure~1.
}
\label{fig3}
\end{figure*}

\begin{figure*}[t]
\includegraphics[bb=10 160 550 640,width=15cm,clip]{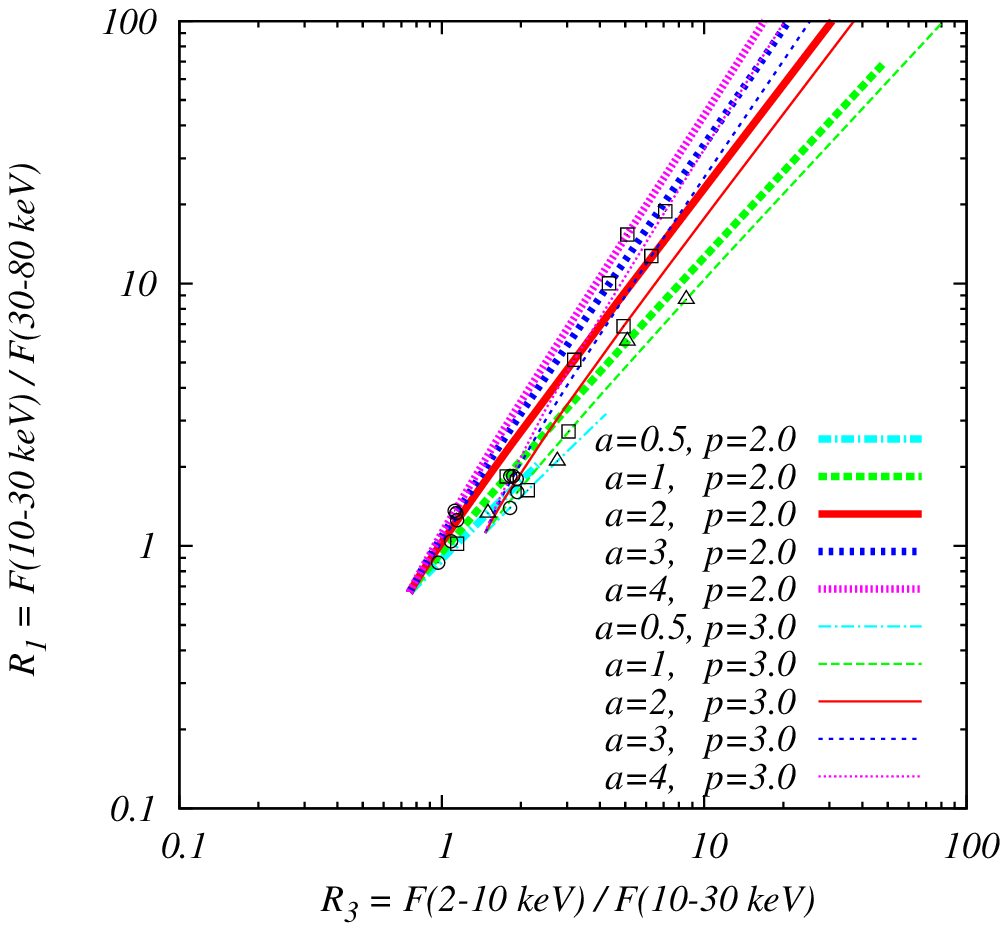}
\caption{The flux ratio
$R_1=F(10-30~{\rm keV})/F(30-80~{\rm keV})$
as a function of $R_3=F(2-10~{\rm keV})/F(10-30~{\rm keV})$.
The meanings of each line and point are the same as in Figure~1.
}
\label{fig4}
\end{figure*}

Figure~\ref{fig3} shows the flux ratio
$R_2=F(2-10~{\rm keV})/F(10-80~{\rm keV})$ as a function of
the 2--10~keV photon index.
In the limit of $BE_{\rm max,e}{}^2\rightarrow\infty$, we have 
$\Gamma_\infty=1.5$ and $R_2\rightarrow0.302$ for $p=2.0$,
but $\Gamma_\infty=2.0$ and $R_2\rightarrow0.774$ for $p=3.0$.

Figure~\ref{fig4} shows 
the flux ratio $R_1=F(10-30~{\rm keV})/F(30-80~{\rm keV})$
as a function of $R_3=F(2-10~{\rm keV})/F(10-30~{\rm keV})$.
In the limit of $BE_{\rm max,e}{}^2\rightarrow\infty$, we have 
$R_1\rightarrow0.668$ and $R_3\rightarrow0.755$ for $p=2.0$,
but $R_1\rightarrow1.12$ and $R_3\rightarrow1.46$ for $p=3.0$.

\begin{figure*}[t]
\includegraphics[bb=50 160 590 640,width=15cm,clip]{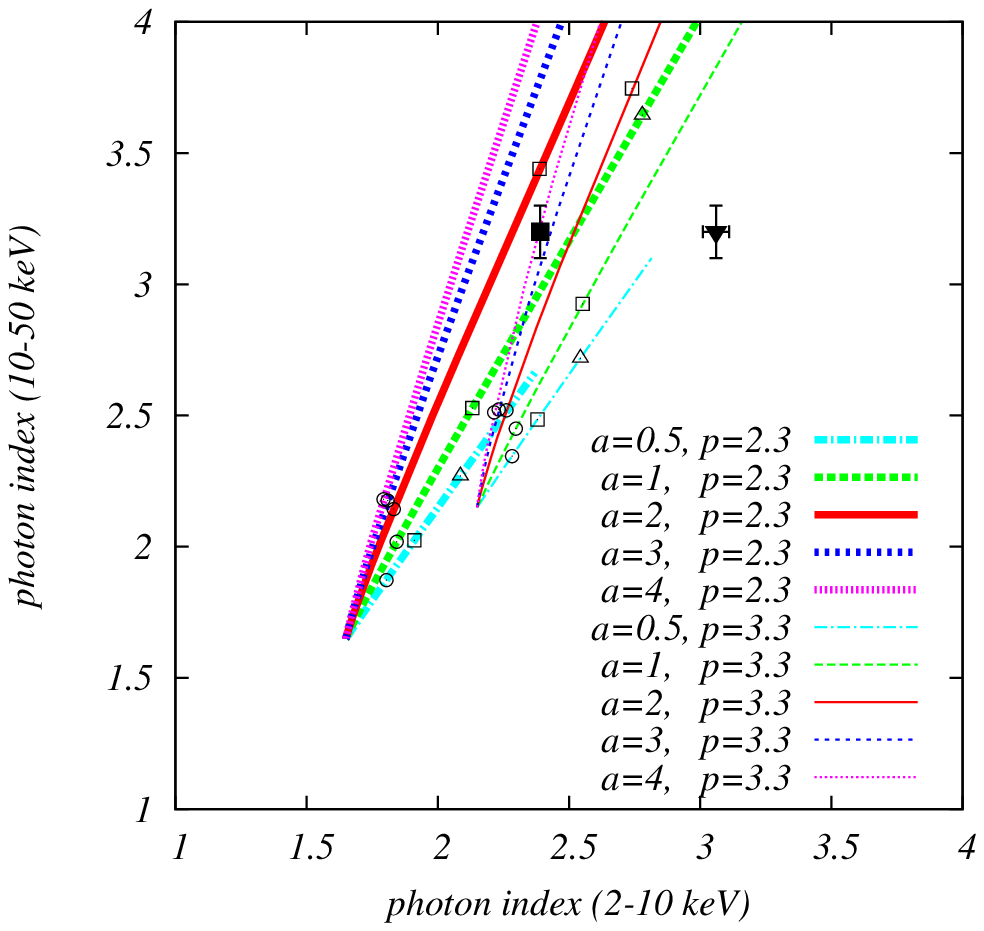}
\caption{
The same as in Fig.~\ref{fig1} but for $p=2.3$ and 3.3.
}
\label{fig5}
\end{figure*}

\begin{figure*}[t]
\includegraphics[bb=10 160 550 640,width=15cm,clip]{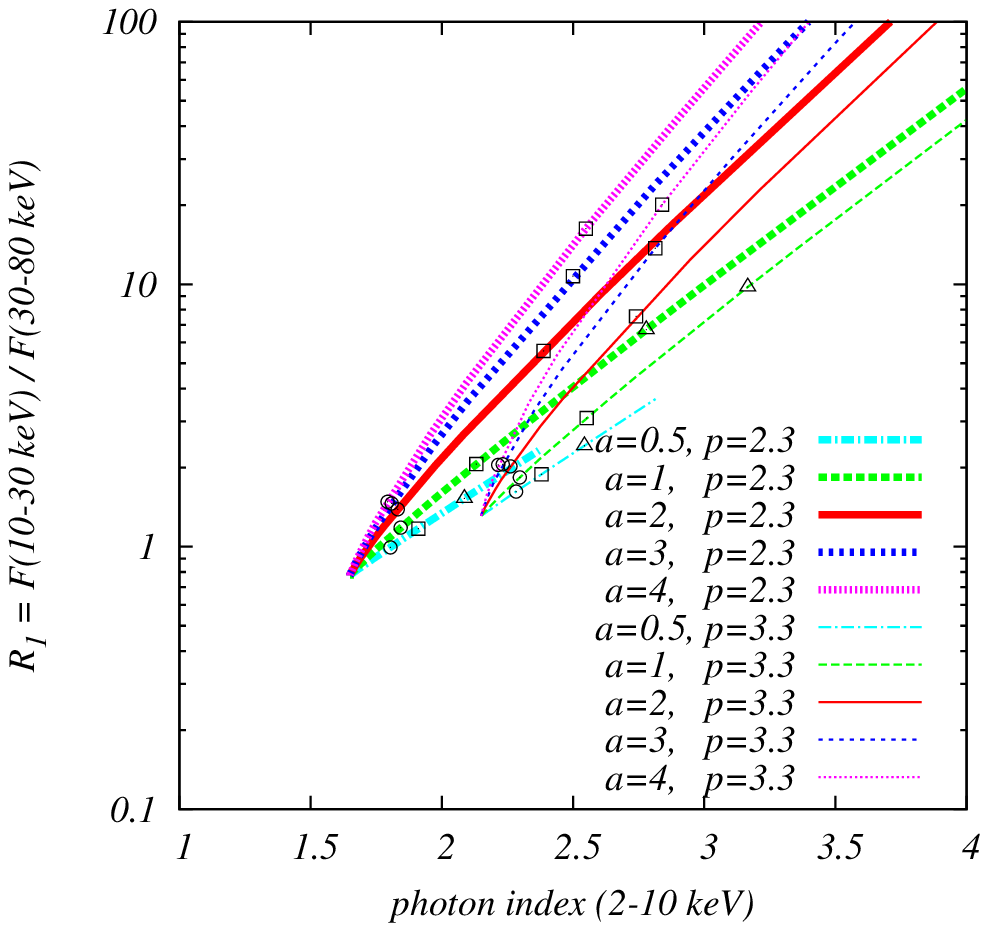}
\caption{
The same as in Fig.~\ref{fig2} but for $p=2.3$ and 3.3.
}
\label{fig6}
\end{figure*}

Figures~\ref{fig5} and \ref{fig6} are the same as Figs.~\ref{fig1}
and \ref{fig2}, respectively, but for $p=2.3$ and 3.3.
The former is typical for the source spectrum of
Galactic cosmic rays,
which is inferred by the propagation model 
(e.g., Strong \& Moskalenko~\cite{strong98}; 
Putze et al.~\cite{putze09}; Shibata et al.~\cite{shibata11}).
It is also expected from gamma-ray observations that 
young SNRs such as Cas~A and Tycho have an energy
spectrum with $p=2.3$
(see \S~\ref{subsec:casA}; Abdo et al.~\cite{abdo10casA}).
The latter case ($p=3.3$) is realized 
if the synchrotron cooling is significant enough to
provide a cooling break below which the electron spectrum
has $p=2.3$.
In the limit of $BE_{\rm max,e}{}^2\rightarrow\infty$, we have 
$\Gamma_\infty=1.65$ and $R_1\rightarrow0.779$ for $p=2.3$,
but $\Gamma_\infty=2.15$ and $R_1\rightarrow1.31$ for $p=3.3$.
Since the differences are too small, we do not show the counterparts of
Figs.~\ref{fig3} and \ref{fig4} for $p=2.3$ and 3.3.

\section{Summary and Discussion}
\label{sec:discuss}

We have derived 
the $\beta$-dependence on $a$ both in the cooling-limited and escape-limited cases 
although based on the test-particle approximation for stationary shocks.
We claim that the cutoff shape of the particle distribution
and/or radiation spectrum potentially becomes a new tool to study
the acceleration mechanism.
Then, we have calculated the spectral properties of synchrotron radiation 
for an electron distribution given by Equation~(1) 
in terms of the flux ratios and photon indices in
various energy bands.
We have presented typical reference lines predicted 
by the simplest one-zone, synchrotron model. 
All the figures in our paper can be directly compared with observations
in the near future.
In the next era of hard X-ray observation with {\it NuSTAR} and {\it ASTRO-H}, 
the observed data should first be compared with those lines.
If the data deviate from the prediction of the model, it
will provide us information on the acceleration mechanism, radiation mechanism,
 the inhomogeneity of the source (validity of one-zone model), and so on.
Note that if we consider the softer band (2--10~keV) in addition to the harder bands
($>10$~keV), then
the energy range becomes broader, so that we can easily constrain
spectral parameters.

The value of $p$ may be determined by the slope of the 
radio synchrotron and/or GeV-to-TeV gamma-ray spectrum.
Once $p$ is independently determined, then comparing
theoretical lines of Figs.~\ref{fig1}, \ref{fig2}, \ref{fig5} 
or \ref{fig6} with observed data, 
which will be provided by {\it NuSTAR} and/or {\it ASTRO-H},
we can constrain the value of $a$.
As discussed in \S~\ref{sec:cutoff_shape},
it is helpful  if the acceleration is
age-limited, cooling-limited or escape-limited.
In particular, we can check if $a\approx2$ or not,
which corresponds to the popular case of Bohm diffusion in
age- or cooling-limited acceleration.
One may expect that $a=2$ is typical.
Indeed, some observational results are consistent with the
case of $a=2$ 
(e.g., Nakamura et al.~\cite{nakamura12}),
although no firm conclusion has been derived.

In addition to other observational information such as a
wide-band radiation spectrum, spatial and temporal variability 
in X-ray and so on,
the value of $a$ will provide another
independent constraint on the magnetic field strength
(for a typical example, see \S~\ref{subsec:rxj1713}),
which, however, is not always applicable.
From another point of view, 
if we know, by other observations, the field strength and understand
how $E_{\rm max,e}$ is determined,
then with the aid of Eq.~(\ref{eq:shape_cooling}), (\ref{eq:shape_age}) 
or (\ref{eq:shape_escape}),
we can further constrain the value of $\beta$,
which gives us rich information on the acceleration process,
in particular, the electron transport mechanism.

Note that our present method is able to constrain the value of $a$
{\it without} detailed spectral fitting which needs brightness.
It is easier than  the
spectral fitting analysis to observationally determine the
flux ratios and/or photon index. 
Even for sources whose X-ray brightness is too small to
perform a precise spectral fitting,
we will be able to discuss the spectral parameter $a$.
Hence we can obtain a larger sample, which enables us to do
the statistical analysis.
Consequently, we will be able to extract more general properties
of particle acceleration at SNRs,
which cannot be done by individual analysis for the small number of bright sources.

Figure~\ref{fig3} shows that all the curves are degenerate
with each other, along which the observed data points should lie
if the X-ray emission arises from synchrotron radiation.
If there are outliers, we can discuss the existence of extra
components, such as a bump in the electron spectrum near $E_{\rm max,e}$
due to effects of pile-up via synchrotron cooling 
(Longair~\cite{longair94}; Drury et al.~\cite{drury99};
Zirakashvili \& Aharonian~\cite{zirakashvili07})
and/or nonlinear acceleration 
(e.g., Malkov \& Drury~\cite{malkov01}),
jitter or diffusive synchrotron emission 
(Toptygin \& Fleishman~\cite{toptygin87}; Medvedev~\cite{medvedev00};
Reville \& Kirk~\cite{reville10}; Teraki \& Takahara~\cite{teraki11}), 
secondary synchrotron radiation that originates from accelerated protons generating
charged pions 
(e.g., Yamazaki et al.~\cite{yamazaki06}),
nonthermal bremsstrahlung emission 
(Laming~\cite{laming01a}, \cite{laming01b}; Vink \& Laming~\cite{vink03};
Vink~\cite{vink08}; Ohira et al.~\cite{ohira12a}), and so on.
Similar discussion may be done with  Fig.~\ref{fig4}.

A caveat is that the one-zone model is adopted for simplicity.
Clearly our plots change if we see emissions from different regions
which have a different parameter set of $B$, $E_{\rm max,e}$, $p$, $a$
as well as flux normalization.
In order to avoid or minimize this problem, 
hard X-ray observations of small emission regions are necessary.
In this sense, the hard X-ray imaging system onboard {\it NuSTAR} or 
{\it ASTRO-H} is useful.

In this paper, we have focused on hard X-rays which arise from synchrotron
radiation of electrons near $E_{\rm max,e}$.
It may be possible to do a similar analysis in gamma-ray bands
using next generation
gamma-ray telescopes like CTA (Actis et al.~\cite{actis11}).
However, at present there are several uncertainties.
First, the emission mechanism is uncertain:
even for the representative case of SNR RX~J1713.7$-$3946,
we have not yet determined whether the gamma-ray emission 
is leptonic or hadronic (see \S~\ref{subsec:rxj1713}).
Second, in the leptonic case, where the gamma-rays arise
from inverse Compton emission of accelerated electrons,
seed photons are uncertain; 
in addition to CMB, infrared and optical photons,
whose intensity is uncertain and depends on the position of
SNRs in the Galactic plane,
significantly contribute to the gamma-ray emission.
Finally, systematic error of the gamma-ray measurement might be too large
for our discussion.
If all of the above uncertainties are removed,
the flux ratio and/or photon index in gamma-ray bands will be another
independent diagnostic. In particular, if gamma-rays are hadronic,
we can obtain the cutoff shape of protons near $E_{\rm max,p}$,
which cannot be done by hard X-ray analysis.
However, even in this case, we need to do a careful analysis because
spatial inhomogeneity of target matter modifies the gamma-ray spectrum 
(e.g., Ohira et al.~\cite{ohira11}; Inoue et al.~\cite{inoue12}).

\subsection{Application to specific cases}

In the following, let us consider the case of Cas~A
and RX~J1713.7$-$3946 as representative examples.

\subsubsection{RX~J1713.7$-$3946}
\label{subsec:rxj1713}

Tanaka et al.~(\cite{tanaka08}) presented results from {\it Suzaku} observations of
RX~J1713.7$-$3946. XIS and HXD, onboard {\it Suzaku}, detected X-rays from this SNR 
in 0.4--12~keV and 12--40~keV bands, and 
measured photon indices in these bands to be $2.39\pm0.01$ and $3.2\pm0.1$, respectively.
We adopt these values as 2--10~keV and 10--50~keV photon indices, respectively. 
Furthermore, using the results of spectral analysis from
Tanaka et al.~(\cite{tanaka08}),
we calculate the energy flux of the whole SNR as
$F(2-10~{\rm keV})=(4.59\pm0.04)\times10^{-10}$erg~s$^{-1}$cm$^{-2}$
and
$F(10-80~{\rm keV})=(1.32\pm0.05)\times10^{-10}$erg~s$^{-1}$cm$^{-2}$,
which lead to the flux ratio,
$R_2 = 3.5\pm0.2$.
The derived value of $R_2$ and the adopted 2--10~keV photon index
are on theoretical lines in  Fig.~\ref{fig3}, 
which implies that
the X-ray emission is the synchrotron radiation.

Recently {\it Fermi} measured the gamma-ray spectrum of
RX~J1713.7$-$3946 in the 3--300~GeV
energy range, whose photon index ($1.5\pm0.1$) is typical for
leptonic inverse Compton emission  from
high-energy electrons with the spectral index $p\approx2.0$ 
(Abdo et al.~\cite{abdo11rxj1713}).
Hence, at first, let us consider 
the possibility that gamma-rays are emitted by inverse Compton scattering.
Assuming CMB and infrared photons are the seeds of the scattering,
Li et al.~(\cite{li11}) extracted the electron distribution from the 
observed gamma-ray spectrum and obtained 
$p\approx2.0$ and $a\approx 0.6$, although the uncertainty is large.
Lee et al.~(\cite{lee12}) set $a=0.5$ to explain the gamma-ray spectrum.
The value $a\approx0.6$ implies $\beta=0.3$ in the age-limited case,
but $\beta=-0.4$ in the cooling-limited case.
The latter is unlikely because the negative value of $\beta$ is
quite unnatural in the context of  magnetic turbulence.
This is also inferred from the fact that the flux ratio of
synchrotron X-rays to inverse Compton gamma-rays leads to the field strength
$B\sim10$--20~$\mu$G 
(e.g., Katz \& Waxman~\cite{katz08}; Yamazaki et al.~\cite{yamazaki09};
Ellison et al.~\cite{ellison10}), so that the effect of
synchrotron cooling is not significant\footnote{
There are several possibilities to explain the
observed time variability and thin filaments of the synchrotron X-rays
without amplification of the magnetic field
(e.g., Katz \& Waxman~\cite{katz08}; Bykov et al.~\cite{bykov08}).
}.
However, one can find from Figs.~\ref{fig1} and \ref{fig5} that
no lines for $a=0.5$ are consistent with the measured photon indices in
the 2--10~keV and 10--50~keV bands. In particular, $a\approx1.5$ if $p\approx2.0$.
This fact was indicated by Tanaka et al.~(\cite{tanaka08}), in which
they found that the X-ray spectrum taken by {\it Suzaku}
is consistent with both cases $a=1$ and $a=2$.
Therefore, the simple leptonic inverse Compton model fails to 
simultaneously explain both the X-ray and the gamma-ray spectral shapes.

On the other hand, hadronic scenarios for the observed hard gamma-ray spectrum
are still viable if we consider the shock-cloud interaction 
(Inoue et al.~\cite{inoue12})
or extreme limit of nonlinear particle acceleration 
(e.g., Yamazaki et al.~\cite{yamazaki09}). 
In these cases, a strong magnetic field ($B\sim0.1$--1~mG) is predicted,
which is also
inferred from the detection of time variability and thin filaments of the
synchrotron X-rays,
although there are some counter arguments 
(e.g., Katz \& Waxman~\cite{katz08}; Bykov et al.~\cite{bykov08};
Reynolds et al.~\cite{reynolds12}).
Hence, the electron acceleration is limited by synchrotron cooling, so that
the spectral index of electrons near $E_{\rm max,e}$ should be 
$p\geq 3.0$ (see \S~\ref{sec:index}).
One can see from Fig.~\ref{fig1} that
if $p\approx3.0$, then the observed photon indices in
the 2--10~keV and 10--50~keV bands tell us $a\geq2$,
so that $\beta\geq1$.
Therefore, Bohm diffusion is consistent with observations.

Note that, however, this argument is not conclusive.
We consider the spectrum of the whole SNR.
As discussed previously, a one-zone model may  not be adequate.
Further observations resolving smaller emission regions 
are necessary.

\subsubsection{Cas~A}
\label{subsec:casA}

XIS and HXD onboard {\it Suzaku} have measured
the spectrum in the 3.4--40~keV band (Maeda et al.~\cite{maeda09}).
They found that the non-thermal component in this energy range
is described by
a single power-law form with a photon index of $3.06\pm0.05$.
Here we adopt this value as the 2--10~keV photon index, although
a thermal component prevents us from precisely determining the index.
Using their fitting parameter,
we have
$F(2-10~{\rm keV})=(6.5\pm0.9)\times10^{-10}$erg~s$^{-1}$cm$^{-2}$.
Other observations in higher energy bands such as
{\it Swift}/BAT (14--195~keV) 
(Baumgartner et al.~\cite{baumgartner12})
and {\it BeppoSAX}/PDS (15--300~keV) (Vink \& Laming~\cite{vink03}), show
a softer photon index of $3.26\pm0.09$ and $3.32\pm0.05$,
respectively. Hence we expect the photon index in the 10--50~keV band
is slightly softer (3.1--3.3) than the value 3.06 measured by 
{\it Suzaku}. Using the parameters given by {\it Swift}/BAT,
we have
$F(10-80~{\rm keV})=(9.6\pm0.7)\times10^{-11}$erg~s$^{-1}$cm$^{-2}$.
Then we obtain the flux ratio $R_2\approx6.8\pm1.6$.

Although the index of the electron spectrum, $p$, is not yet fixed,
the measured gamma-ray spectrum shows that $p$ is larger than 2.0 
(Abdo et al.~\cite{abdo10casA}).
Here we adopt $p\approx2.3$ as a typical value.
Then, one can plot the observed data in Figures~\ref{fig3} and \ref{fig5}.
Placing the observed 2--10~keV photon index of $3.06\pm0.05$ and
the flux ratio $R_2\approx6.8\pm1.6$ in Fig.~\ref{fig3},
we find that the data are marginally consistent with the theoretical
prediction for the synchrotron radiation. However, 
the observed data point of 2--10~keV and 10--50~keV photon indices
does not lie on any lines in Fig.~\ref{fig5}.
Furthermore, the observed data show that for a given value 
of 2--10~keV photon index,
the 10--50~keV photon index is smaller than expected via
the synchrotron radiation.
Hence, one can claim the existence of an extra component.
At present, the origin of the hard X-rays above 10~keV for Cas~A is a matter
of debate. In particular we are interested in whether the emission is 
nonthermal bremsstrahlung or synchrotron
(Laming~\cite{laming01a}, \cite{laming01b}; Vink \& Laming~\cite{vink03};
Vink~\cite{vink08}).
Our discussion here may contribute to this topic.
Note that, however, this argument is based on a one-zone model.
At present, it is uncertain whether 2--10~keV and 10--80~keV
X-rays come from the same emission region, which is left for
future observations.

One may insist on explaining the observed photon indices
by only the synchrotron radiation.
Indeed, it is marginally possible that the observed 2--10~keV
and 10--50~keV photon indices lie on the theoretical line of 
$p\geq3$ and $a\approx0.5$. In this case, the acceleration is
not limited by synchrotron cooling because generally $\beta>0$ is
expected (see Eq.~\ref{eq:shape_cooling}).
This is apparently inconsistent with the observational implications
that magnetic field strength is large enough ($B\geq0.5$~mG)
for the electron acceleration to be cooling limited
(Vink \& Laming~\cite{vink03}; Bamba et al.~\cite{bamba05a}).

\begin{acknowledgements}
We would like to thank 
Tsunehiko~Kato, Kohta~Murase,
Takanori~Sakamoto, Atsumasa~Yoshida, Tsuyoshi~Inoue and
Tohru~Shibata, and the anonymous referee
for useful comments.
This work was supported in part 
by the fund from Research Institute, Aoyama Gakuin University
(R.~Y. and A.~B.), and by grant-in-aid from the 
Ministry of Education, Culture, Sports, Science,
and Technology (MEXT) of Japan,
No.~24$\cdot$8344 (Y.~O.), No.~24840036 (M.~S.) and
No.~22684012 (A.~B.)
\end{acknowledgements}

\appendix                  

\section{Characteristic energies}
\label{sec:appendix}

In this section, we assume the Bohm diffusion, $K(E)\propto E$, for simplicity.
More detailed analysis is found in Ohira et al.~(\cite{ohira12}).
The extension to a more general case ($K(E)\propto E^\beta$) is easy and omitted here.

\subsection{Maximum energy of accelerated electrons and protons}
\label{subsec:Emax}

Let $v_{\rm s}$ and $B_{\rm d}$ 
be the shock velocity and the downstream magnetic field, respectively.
First, suppose that the maximum energy of electrons
is determined from the balance of the synchrotron loss and acceleration.
Then, equating the acceleration time 
$t_{\rm acc}(E)=20\xi cE/3ev_{\rm s}^2B_{\rm d}$
with the synchrotron cooling time, 
$t_{\rm syn}(E) = 125~{\rm yr}(E/10~{\rm TeV})^{-1}(B_{\rm d}/100~\mu{\rm G})^{-2}$, 
we obtain
\begin{equation}
E_{\rm max}^{\rm (cool)} = \frac{24}{\xi^{1/2}}
\left(\frac{v_{\rm s}}{10^8{\rm cm}~{\rm s}^{-1}}\right)
\left(\frac{B_{\rm d}}{10~\mu{\rm G}}\right)^{-1/2} ~{\rm TeV}~~,
\label{eq:Emax_loss}
\end{equation}
where $\xi$ is a gyro factor.
On the other hand, if the cooling is not significant,
that is, $t_{\rm acc}(E)$, $t_{\rm age}\ll t_{\rm syn}(E)$, then
the maximum energy can be determined by the condition 
$t_{\rm acc}(E) = t_{\rm age}$, and we obtain
\begin{equation}
E_{\rm max}^{\rm (age)}= \frac{4.8\times10^2}{\xi}
\left(\frac{v_{\rm s}}{10^9{\rm cm}~{\rm s}^{-1}}\right)^2
\left(\frac{B_{\rm d}}{10~\mu{\rm G}}\right)
\left(\frac{t_{\rm age}}{10^3{\rm yr}}\right) ~{\rm TeV}~~.
\label{eq:Emax_age}
\end{equation}
For young SNRs, we typically expect
the maximum energies of electrons and protons, $E_{\rm max,e}$
and $E_{\rm max,p}$ are $E_{\rm max}^{\rm (cool)}$ and
$E_{\rm max}^{\rm (age)}$, respectively.
Then, using equation~(\ref{eq:Emax_loss}) and (\ref{eq:Emax_age}), 
we eliminate $v_{\rm s}^2/\xi$ and obtain
\begin{equation}
E_{\rm max,p}\approx 83
\left(\frac{E_{\rm max,e}}{10~{\rm TeV}}\right)^2
\left(\frac{B_{\rm d}}{100~\mu{\rm G}}\right)^2
\left(\frac{t_{\rm age}}{10^3{\rm yr}}\right) ~{\rm TeV}~~.
\label{eq:Emax_age2}
\end{equation}

\subsection{The cooling break energy}
\label{subsec:cooling}

Let $t$ be the characteristic time of SNR evolution,
which may be the expansion time of $t_{\rm age}$ itself.
Then the cooling break, $E_{\rm b}$, appears in the electron spectrum 
at the energy where $t_{\rm syn}(E_{\rm b})=t$, that is,
\begin{equation}
E_{\rm b} = 12.5~{\rm TeV}
\left(\frac{t}{10^2{\rm yr}}\right)^{-1}
\left(\frac{B_{\rm d}}{100~\mu{\rm G}}\right)^{-2}~~.
\label{eq:coolingbreak}
\end{equation}
The cooling break appears if $E_{\rm b}< E_{\rm max,e}=E_{\rm max}^{\rm (cool)}$,
which can be rewritten as
\begin{equation}
B_{\rm d}>139~\mu{\rm G}~ \xi^{1/3}
\left(\frac{v_{\rm s}}{10^8{\rm cm}~{\rm s}^{-1}}\right)^{-2/3}
\left(\frac{t}{10^2{\rm yr}}\right)^{-2/3}~~.
\label{eq:cond_Eb}
\end{equation}

\section{Analytical derivation of asymptotic spectral shape}

\subsection{The case of cooling-limited acceleration}
\label{sec:appendix2}

Zirakashvili \& Aharonian~(\cite{zirakashvili07})
 obtained the asymptotic electron spectrum 
near the maximum electron energy in the
cooling-dominated phase, such as
\begin{equation}
N(E)= A_0(E)\exp\left[S_0(E)\right] ~~,
\end{equation}
\begin{equation}
A_0(E) = E^{-1/2}\exp\left[
\int^E\frac{\sqrt{K_2}\frac{\partial}{\partial E}\sqrt{b_2}
+\sqrt{K_1}\frac{\partial}{\partial E}\sqrt{b_1}}
{\sqrt{b_1K_1}+\sqrt{b_2K_2}} d E'
\right]~~,
\end{equation}
\begin{equation}
S_0(E) = -\left(\frac{\gamma_s}{v_{\rm s}}\right)^2
\int^E\left(
\frac{\sqrt{b_1K_1}+\sqrt{b_2K_2}}{E'}\right)^2 dE'~~,
\end{equation}
where $\gamma_s=3r/(r-1)$ ($r$ is the shock compression ratio),
and $v_{\rm s}$ is the shock velocity.
Functions $K(E)$ and $b(E)=dE/dt$ are the
diffusion coefficient and energy loss rate of electrons, respectively.
Subscripts 1 and 2 indicate upstream and downstream regions
of the shock, respectively. 
In the case of synchrotron cooling, $b(E)$ is proportional to $E^2$.

We assume that $K(E)=K_0E^\beta$ and $b(E)=b_0E^2$ where $K_0$ and $b_0$ are
constants, and  that
ratios $b_1/b_2$ and $K_1/K_2$ are also constant. Then we obtain
\begin{equation}
A_0(E)\propto E^{1/2}~~.
\label{eq:A0}
\end{equation}
However, for typical parameters of young SNRs, 
this positive slope can hardly be seen in the X-ray emission.
Furthermore, neglecting non-dimensional terms on the order of unity,
we derive $S_0(E)\approx-(K_0b_0/v_{\rm s}^2)E^{\beta+1}$.
By the way, acceleration time and synchrotron cooling time are calculated as
$t_{\rm acc}(E)\approx K_0E^\beta/v_{\rm s}^2$
and $t_{\rm syn}(E)\approx(b_0E)^{-1}$, respectively, where we again
neglect terms on the order of unity.
Then equating them, we obtain
$E_{\rm max}^{\rm (cool)}\approx (v_{\rm s}^2/K_0b_0)^{1/(\beta+1)}$.
Therefore, we finally derive
\begin{equation}
S_0(E)\approx-\left(\frac{E}{E_{\rm max}^{\rm (cool)}}\right)^{\beta+1}~~.
\label{eq:S0}
\end{equation}

\subsection{The case of escape-limited acceleration}
\label{sec:escape}

We assume the test-particle regime and place
a free escape boundary upstream at a distance of $\ell$ away from the shock
front $x=0$,
that is, the particle distribution function is zero at $x=-\ell~(<0)$.  
Then the stationary transport equation is solved to find a
particle spectrum around the shock front given by 
(Caprioli et al.~\cite{caprioli09}; Reville et al.~\cite{reville09})
\begin{equation}
N(E) \propto \exp\left\{
-\frac{3r}{r-1}  \int^{E}
\frac{{\rm d}\,\log E'}
{1-\exp\left[-v_{\rm s}\ell/K(E')\right]}
\right\}~~,
\end{equation}
where $v_{\rm s}$ is the velocity of the shock.
In the escape-limited case, the maximum energy is determined by
the condition (Ohira et al.~\cite{ohira10})
\begin{equation}
\frac{K(E_{\rm max}^{\rm (esc)})}{v_{\rm s}} = \ell~~.
\end{equation}
Then, one can see
$K(E)/{v_{\rm s}}\ell=K(E)/K(E_{\rm max}^{\rm (esc)})=(E/E_{\rm max}^{\rm (esc)})^\beta$.
Changing variables into $y(E')=(E'/E_{\rm max}^{\rm (esc)})^\beta$, we derive
\begin{equation}
N(E) \propto \exp\left\{
-\frac{3r}{r-1}  \times\frac{1}{\beta}
\int^{y(E)}
\frac{{\rm d}\,\log y}{1-e^{-1/y}}
\right\}~~.
\end{equation}
In the case of $E\ll E_{\rm max}^{\rm (esc)}$,
the term $e^{-1/y}$ can be neglected, so that we obtain
$N(E)\propto E^{-3r/(r-1)}$.
On the other hand, if $E\gg E_{\rm max}^{\rm (esc)}$,
we approximate $1-e^{-1/y}\approx1/y$, resulting in
$N(E)\propto\exp[-(E/E_{\rm max}^{\rm (esc)})^\beta]$,
where  a term on the order of unity is again neglected.

\label{lastpage}

\end{document}